\documentclass[]{spie}  %>>> use for US letter paper
\usepackage{amsmath,amsfonts,amssymb}
\usepackage{graphicx}
\usepackage[colorlinks=true, allcolors=blue]{hyperref}
\usepackage{tabularx,booktabs,subfigure,enumerate} %Tsuzuki add

\title{The Infrared Imaging Spectrograph (IRIS) for TMT: optical design of IRIS imager with "Co-axis double TMA"} 

\author[a]{Toshihiro Tsuzuki}
\author[a]{Ryuji Suzuki}
\author[a]{Hiroki Harakawa}
\author[a]{Bungo Ikenoue}
\author[b]{James Larkin}
\author[c]{Anna Moore}
\author[a]{Yoshiyuki Obuchi}
\author[d]{Andrew C Phillips}
\author[a]{Sakae Saito}
\author[a]{Fumihiro Uraguchi}
\author[c]{James Wincentsen}
\author[e]{Shelley Wright}
\author[a]{Yutaka Hayano}

\affil[a]{Advanced Technology Center, National Astronomical Observatory of Japan, 2-21-1 Osawa, Mitaka, Tokyo, Japan}
\affil[b]{Department of Physics and Astronomy, Univ. of California, Los Angeles, CA 90095-1547}
\affil[c]{Caltech Optical Observatories,1200 E California Blvd., Pasadena, CA 91125}
\affil[d]{University of California Observatories, CfAO, University of California, 1156 High St., Santa Cruz, CA 95064}
\affil[e]{Center for Astrophysics and Space Sciences, Univ. of California, San Diego, La Jolla, CA 92093}

\authorinfo{Further author information: (Send correspondence to Toshihiro Tsuzuki)\\Toshihiro Tsuzuki: E-mail: toshihiro.tsuzuki@nao.ac.jp, Telephone: +81-422-34-3891}

\begin{document} 
\maketitle

\begin{abstract}
IRIS (InfraRed Imaging Spectrograph) is one of the first-generation instruments for the Thirty Meter Telescope (TMT). IRIS is composed of a combination of near-infrared (0.84--2.4 $\mu$m) diffraction limited imager and integral field spectrograph. 
To achieve near-diffraction limited resolutions in the near-infrared wavelength region, IRIS uses the advanced adaptive optics system NFIRAOS (Narrow Field Infrared Adaptive Optics System) and 
integrated on-instrument wavefront sensors (OIWFS). 
However, IRIS itself has challenging specifications. First, the overall system wavefront error should be less than 40 nm in Y, z, J, and H-band and 42 nm in K-band over a 34.0 $\times$ 34.0 arcsecond field of view. 
Second, the throughput of the imager components should be more than 42 percent. To achieve the extremely low wavefront error and high throughput, all reflective design has been newly proposed. 
We have adopted a new design policy called "Co-Axis double-TMA", which cancels the asymmetric aberrations generated by "collimator/TMA" and "camera/TMA" efficiently. 
The latest imager design meets all specifications, and, in particular, the wavefront error is less than 17.3 nm and throughput is more than 50.8 percent. 
However, to meet the specification of wavefront error and throughput as built performance, the IRIS imager requires both mirrors with low surface irregularity after high-reflection coating in cryogenic 
and high-level Assembly Integration and Verification (AIV). 
To deal with these technical challenges, we have done the tolerance analysis and found that total pass rate is almost 99 percent in the case of gauss distribution and more than 90 percent in the case of 
parabolic distribution using four compensators. We also have made an AIV plan and feasibility check of the optical elements. In this paper, we will present the details of this optical system.  
\end{abstract}


\section{INTRODUCTION}
\label{sec:INTRODUCTION}  % \label{} allows reference to this section

IRIS (InfraRed Imaging Spectrograph \cite{larkin2010infrared, moore2014infrared}) is one of the first-generation instruments for the Thirty Meter Telescope (TMT \cite{sanders2013thirty}). 
IRIS is a fully cryogenic instrument that combines a "wide-field” diffraction limited imager and an optically following integral field spectrograph both covering wavelengths from 0.84 to 2.4 $\mu$m. 
To achieve near-diffraction limited resolutions in the near-infrared wavelength region, IRIS can utilize the advanced adaptive optics system NFIRAOS (Narrow Field Infrared Adaptive Optics System \cite{herriot2012tmt}) and 
also integrated on-instrument wavefront sensors (OIWFS \cite{dunn2014instrument}). 
 
One of the most challenging tasks of the IRIS imager is to achieve extremely low wavefront error (less than 40 nm in Y, z, J, and H-band and 42 nm in K-band) in as-built performance .
To deal with the problem, the optical design started with an all-refractive design using apochromatic triplets (ApTs) called "ApT/ApT" design as a product of the conceptual design phase study.
The demands for higher throughput and wider Field of View (FoV) changed the "ApT/ApT" design to refractive-reflective design using the ApT as a collimator and the three-mirror anastigmat (TMA) as a camera
 (called "ApT/TMA" design).
However, an all-reflective design with potentially the highest throughput has not been achieved.

In this paper, we describe the details about our newly proposed all-reflective design and also the new design method called "Co-axis double TMA".
Section 2 gives the specifications of IRIS imager. 
Section 3 covers optical design policy and section 4 covers the optical design method called "Co-axis double TMA". 
Section 5 gives the nominal performances and as-built performances of the design.
Finally, In Section 6, an AIV plan is proposed. 

\section{SPECIFICATIONS}
\label{sec:SPECIFICATIONS}

\begin{figure}[htbp]
	\begin{center}
	\begin{tabular}{cc}
	   \subfigure[Detector size and format]{
			\includegraphics[width=0.45\textwidth]{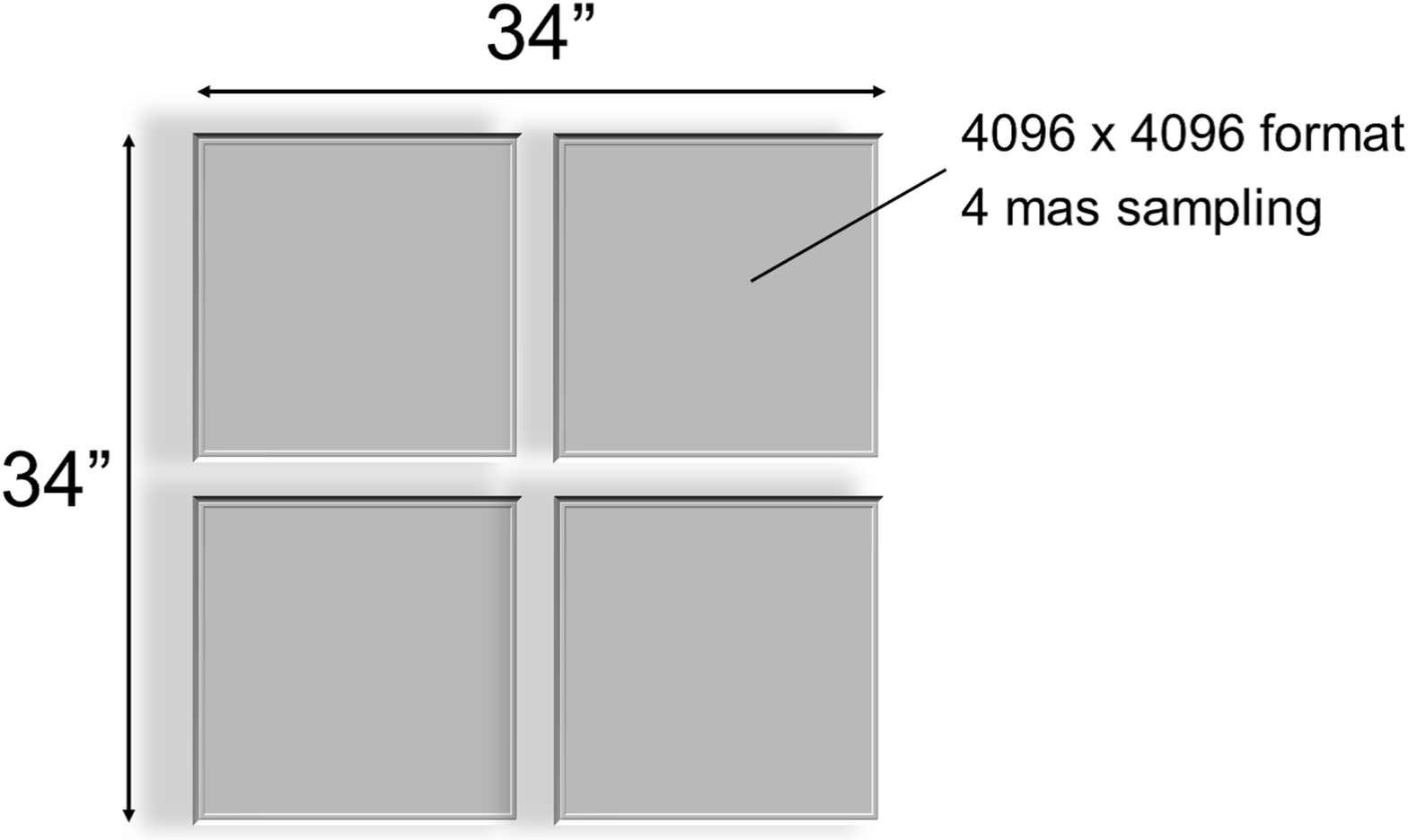}
			} &
	   \subfigure[Detector arrangement]{
			\includegraphics[width=0.45\textwidth]{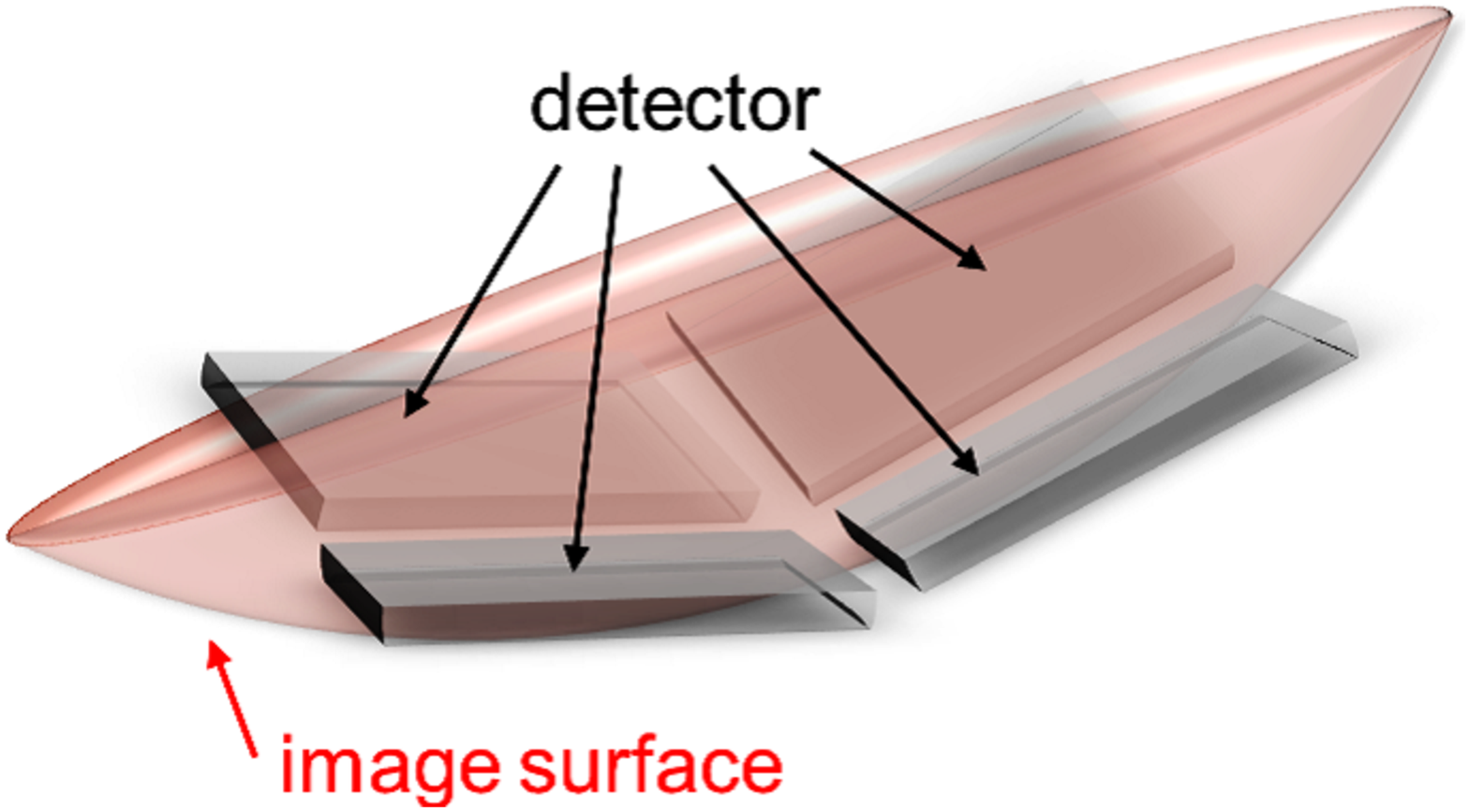}
			}
	\end{tabular}
	\end{center}
\caption{Conceptual images of the detectors : (a) detector size and format, (b) detector arrangement. Four flat detectors with 4096 $\times$ 4096 format with 4 mas sampling can be tilted independently 
			so that they can fit the image surface.}
\label{fig:IRIS_SciencePath_DetectorInfomation}
\end{figure}

Table \ref{tab:IRIS_SciencePath_DesignSpecification} shows the major specifications for the imager optical design. 
These specifications are determined by the upper-level  subsystem requirements called level 3.
Regarding detectors, detectors are assumed to be four Hawaii 4RG (Each detector has 4096 $\times$ 4096 format with 4 mas sampling, See Figure \ref{fig:IRIS_SciencePath_DetectorInfomation}(a)) 
The Four flat detectors can be tilted independently so that they fit the image surface (See Figure \ref{fig:IRIS_SciencePath_DetectorInfomation}(b)).

\vspace{0.01\textwidth}
%-------------
\begin{table}[htbp]
	\begin{center}
		\begin{tabular}{lp{11.5cm}}
			\toprule[1pt]
					\textbf{Items}& \textbf{Specifications} \\\hline\hline
					Wavelength range & 0.84 -- 2.4 microns  \\
					Field of view & 34.0$\times$34.0 arcsecond \\
					Configuration & Collimator (F/15) + camera (F/17.19 +/- 0.02) \\
					Wavefront error & WFE should be less than 40 nm in Y, z, J, and H-band and 42 nm in K-band after 60$\times$60 order correction by NFIRAOS DMs. \\
					Throughput & The throughput of the imager components should be more than 42 \% \\
					Ghost & (Discussions on-going with the science team.) \\
					Distortion & The distortion should be less than one percent \\
					Size & The imager should fit into the dewar (Dewar size are shown in Figure \ref{fig:IRIS_SciencePath_DetectorInfomation}(a)) \\
					Pupil image & The collimator optics should create a real image of the pupil in the collimated beam. \\
					Pupil aberration & Geometrical spot diameter should be less than 0.4 \% of the diameter of the pupil image \\
					Fore optics & NFIRAOS science path \\
					Detector & Detectors are assumed to be four Hawaii 4RG (Each detector has 4096 $\times$ 4096 format with 4 mas sampling, See Figure \ref{fig:IRIS_SciencePath_DetectorInfomation}(a))  \\
					Image surface & Four flat detectors that can be tilted independently (See Figure \ref{fig:IRIS_SciencePath_DetectorInfomation}(b)) \\
					Collimated beam & The collimator should provide a collimated beam of enough length after cold stop for ADC \& filter wheel.  \\
			\bottomrule[1pt]
		\end{tabular}
	\vspace{0.02\textwidth}
	\caption{Major Requirements on the IRIS imager} 
	\label{tab:IRIS_SciencePath_DesignSpecification}
	\end{center}
\end{table}
%-------------

%-------------
\begin{figure}[htbp]
	\begin{center}
		\includegraphics[width=0.90\textwidth]{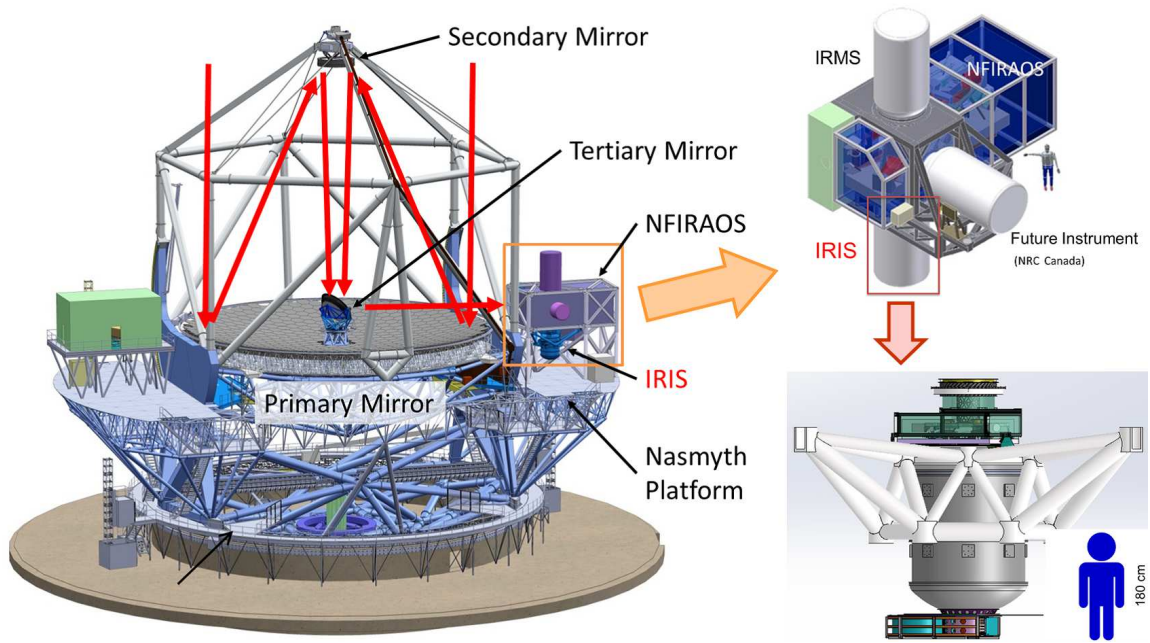}
	\end{center}
\caption{Schematic view of TMT/NFIRAOS/IRIS. The light collected by the TMT passes through NFIRAOS on the Nasmyth platform and finally enters IRIS.}
\label{fig:IRIS_SciencePathDesign_TMT_NFIRAOS_IRIS}
\end{figure} 
%-------------

%-------------
%-------------

\section{OPTICAL DESIGN POLICY}
\label{sec:OPTICAL DESIGN POLICY}
In this section, characteristics of the optical system and the design guideline are described. 

\subsection{Characteristics of the Optical System}
The optical system described in the specification is the imaging optical system equipped with cold stop. 
The minimum power structure to achieve these functions consists of two pieces; the first power (collimator) turns F/15 diverging beam into collimated beam 
and the second power (camera) turns collimated beam into F/17.19 converging beam. 
The collimator’s performance is defined by the spot diameter of the pupil aberration, 
and the final imaging performance with the collimator and the camera is defined by the wavefront aberration at the final image surface. 
Note that due to the difference in F-number between the collimator and the camera (i.e. F/15 and F/17.19,respectively), their powers are different as well. 
In addition, for the collimated light portion, there inserted are the cold stop, ADC, filter wheel and several fold mirrors for pupil viewing optics. 
Therefore, an optical path of at least 600 mm is required for the collimated beam.

\subsection{Throughput}
Regarding the throughput of this optical system, as high as possible a throughput is greatly desired.
To achieve this, the use of mirrors, rather than lenses, is required. 
Also, the number of optical elements needs to be minimized.

\subsection{Nominal Performance}
First of all, this optical system requires extremely small wavefront aberration (less than 42 nm).
To achieve this, the field curvature at the final IRIS imaging surface needs to be made small.
The reason is that the four tilted plane detectors described in Figure \ref{fig:IRIS_SciencePath_DetectorInfomation} cannot completely 
run along the IRIS imaging surface except in the case of a flat imaging surface.
Therefore, if the field curvature at the final IRIS imaging surface is large, 
the non-negligible deterioration of the wavefront aberration caused by the position deference between imaging surface and detector’s light receiving surface occurs.

Next, the correction on the residual aberration of the fore optics (TMT + NFIRAOS) needs to be made.
Regarding the field curvature, the positive field curvature with the radius of curvature at 1398 mm exists at the focal plane of the fore optics (i.e. 
the incident position of the IRIS imager).
This means that even if the effects are minimized by tilting the four plane detectors, the wavefront aberration of 20 nm RMS still remains.
Therefore, the IRIS imager itself needs to have the negative field curvature 
in order to minimize the field curvature at the final IRIS imaging surface (TMT + NFIRAOS + IRIS).

\begin{figure}[htbp]
	\begin{center}
	\begin{tabular}{cc}
	   \subfigure[Residual aberrations at the focal plane of the fore optics (TMT + NIFRAOS)]{
			\includegraphics[width=0.65\textwidth]{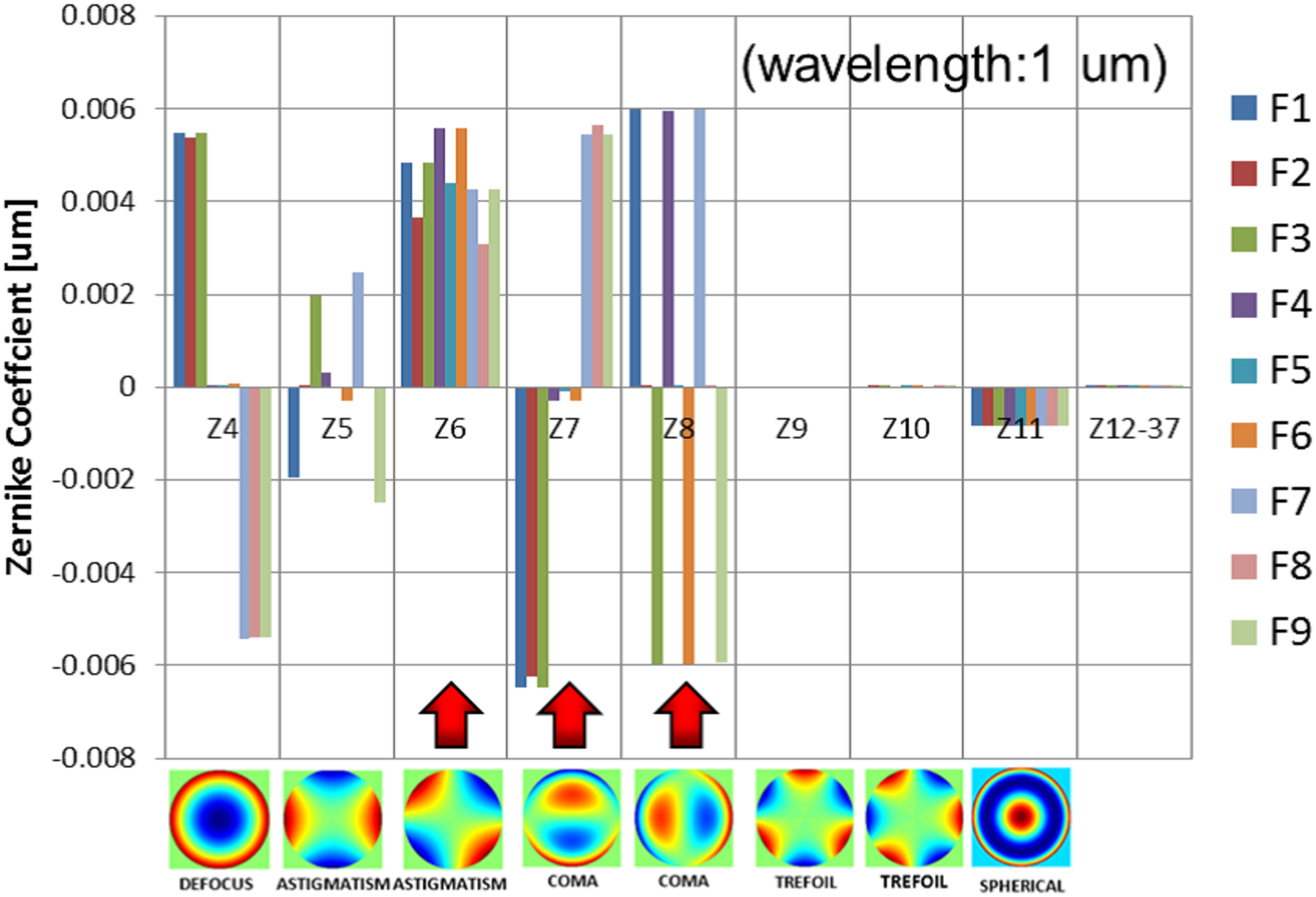}
			} &
	   \subfigure[Definition of the FoV numbers]{
			\includegraphics[width=0.25\textwidth]{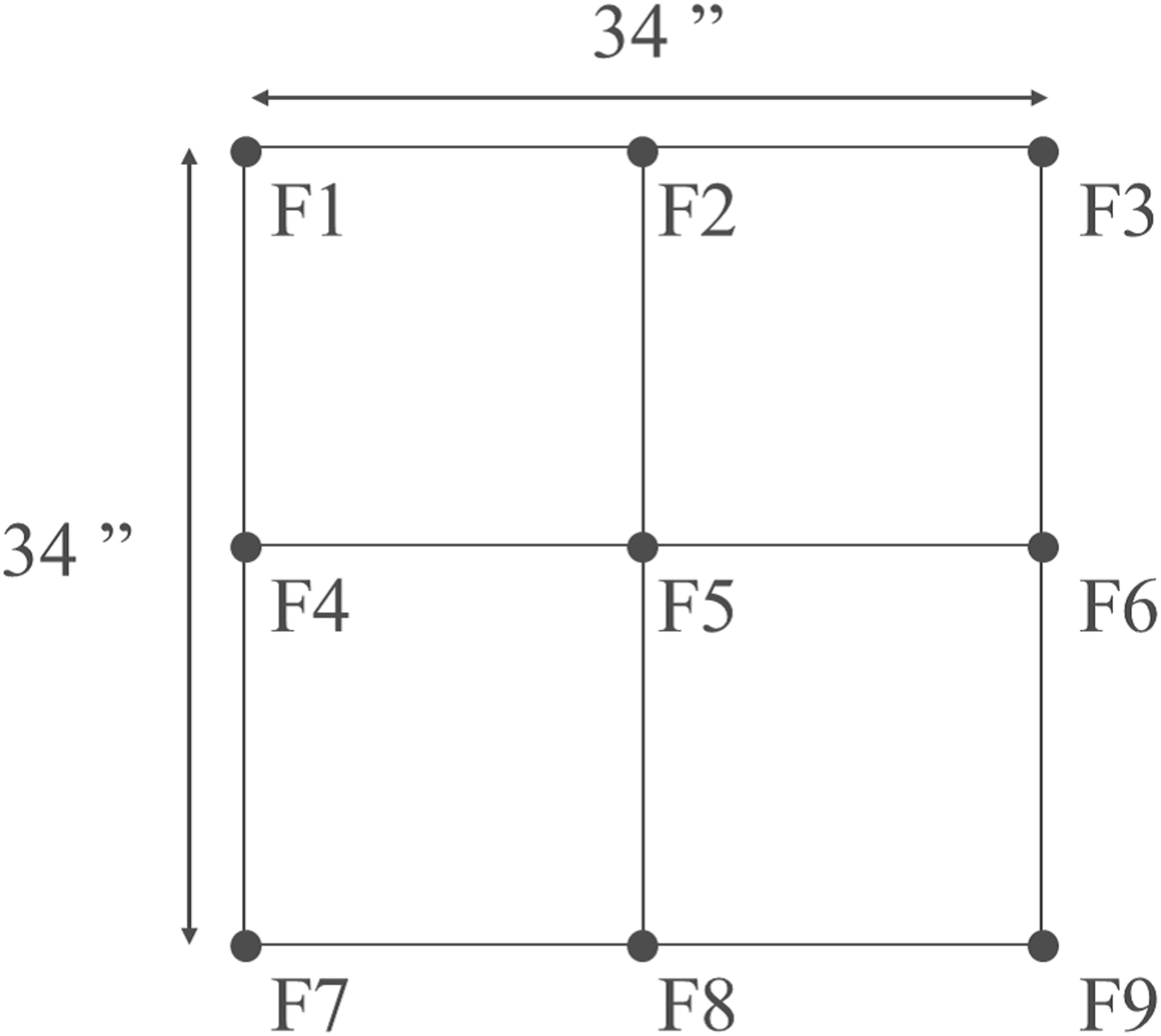}
			}
	\end{tabular}
	\end{center}
\caption{Residual aberrations at the focal plane of the fore optics (TMT + NIFRAOS) (wavelength : 1 $\mu$m, excluded the field curvature).
			Defocus caused by image tilt (Z4), astigmatism Z6) and coma aberration (Z7, Z8) are dominant.
			F1 to F9 indicate the FoV numbers (refer to the Figure on the right for the definition)}
\label{fig:IRIS_SciencePath_TMTNFIRAOS_WavefrontError}
\end{figure}

There exist other aberrations of the fore optics as well.
Figure \ref{fig:IRIS_SciencePath_TMTNFIRAOS_WavefrontError} shows the wavefront aberrations of the fore optics at each FoV.
From this, it is clear that the image tilt, coma and astigmatism aberration are dominant as residual aberrations of fore optics.
Therefore, the IRIS imager needs to correct these rotationally asymmetric aberrations by itself.

The fore optics has AO (Adaptive Optics) and it can correct the wavefront aberration by transforming DMs (Deformable Mirrors).
However, since the DMs are located in the pupil position, it can only correct the wavefront aberration that is common to each FoV.
This means that to maximize the effect of DMs, 
it would be the best if the aberration at the final IRIS imaging surface (TMT + NFIRAOS + IRIS) is rotationally symmetric.
For this reason, it would be preferable for IRIS imager to eliminate the rotationally asymmetric aberration 
caused by fore optics (TMT + NFIRAOS).

%-------------
%-------------

\subsection{As-bulit Performance}
This optical system requires extremely small wavefront aberration (less than 42 nm) after the assembly is completed.
Therefore, the extra wavefront aberration caused by production or assembly needs to be taken into consideration.

Starting with the production, the low surface irregularity of the optical elements and the refractive index homogeneity of the glass material 
would result in the deterioration of the wavefront aberration. 
In the case where the transmission type optical elements such as lenses are used, the effect of the surface irregularity on the wavefront aberration is relatively small 
but the refractive index homogeneity of the glass material has a large impact.
On the other hand, in the case where the reflection type optical elements such as mirrors are used, the surface irregularity has a larger impact on the wavefront aberration 
but the refractive index homogeneity of the mirror substrates does not need to be taken into consideration.
There is a production limitation in the surface irregularity so it would be best to minimize the number of optical elements to be used during the design phase.

Regarding the assembly process, it is important to decrease the tolerance sensitivity.
Since the wavefront aberration required in the specification is the worst acceptable value within all of FoV,
it is preferable for the low-order aberrations at the final IRIS imaging surface (TMT + NFIRAOS + IRIS) to be small in order to decrease the tolerance sensitivity.
The low-order aberrations caused by manufacturing errors and installation errors can be separated into two groups: 
rotationally symmetric aberrations and rotationally asymmetric aberrations, 
but the rotationally symmetric aberrations can be corrected by DMs.
Thus, for the IRIS imaging optical system itself, it would be the best to eliminate rotationally asymmetric aberration caused by the fore optics (TMT + NFIRAOS). 

Also, components (i.e. the collimator block and the camera block ) are to have the almost the same level of tolerance sensitivity.
Furthermore, in the case where the collimator block and the camera block consist of multiple optical elements,
it would be preferable that their tolerance sensitivities are evenly distributed as well.
Considering the compensator, it would be the best that the aberrations produced by one element could correspond to the opposite aberration produced by other element. 

In addition, minimizing the degree of freedom for manufacturing error (i.e. less number of optical elements) is preferred.

\subsection{Performance in Operation}
There are many factors deteriorating the wavefront aberration during the operation of the instrument.
First, there is an issue with the vibration transmitted to the instrument.
When the vibration is transmitted to each optical element, the minute position changes of the elements occur.
The infinitesimal displacement causes the chief ray shift and finally results in the wavefront aberration error.
For aberrations other than the chief ray shift, their effects can be ignored since the vibration is relatively small. 

Next, we need to consider the environmental condition during the operation.
Since IRIS is an infrared imaging instrument, it needs to be operated in a low-temperature environment (77K).
Thus, the whole IRIS optical system is placed in a vacuum chamber. Therefore, the window of the vacuum chamber may change shape due to the pressure difference.

Also, since the system is placed in a low-temperature environment, positions of the optical elements may change due to the thermal contraction of the stage.
In addition, the difference in coefficient of thermal expansion (CTE) between the optical elements and their holding mechanism may cause their surfaces to change.
The former issue can be corrected by adjusting the focus because it corresponds to the multiplication of the optical system by scale,
but the latter issue could become a major factor of the wavefront error since the shape of the optical elements themselves change. 

The other factor is the rotation of the image surface during the operation. 
TMT is the Nasmyth altazimuth mount type reflecting telescope.
Therefore the FoV rotates along with the elevation angle of the telescope.
IRIS is correcting this by rotating itself.
This operation corresponds to the relative rotation along the optical axis between fore optics (TMT + NFIRAOS) and IRIS.
It results in the rotation of the residual aberration of the fore optics (Figure \ref{fig:IRIS_SciencePath_TMTNFIRAOS_WavefrontError}) ,
especially the rotationally asymmetric aberrations.

Last, the rotation of the ADC (Atmospheric Dispersion Compensator) needs to be taken into consideration.
ADC is needed to compensate the atmospheric dispersion effects and IRIS uses the rotational ADC inserted in the collimated light region due to its limited space.
ADC rotation according to the observed zenith angle of the telescope corresponds to the rotation and shift of the optical axis between the collimator block and the camera block,
which causes the coordinate systems of the collimator block and the camera block to relatively shift or tilt. This may deteriorate the imaging performance.

\subsection{Packaging}
As mentioned above, the whole IRIS optical system is placed in the vacuum chamber since it is operated in a low-temperature environment (77K).
Therefore, the IRIS imager needs to be fit within a space about 1500 mm in diameter and 1150 mm in height.
The appearances of the vacuum chamber and the space available for the IRIS imager are shown in the Figure \ref{fig:IRIS_SciencePath_Dewar} below. 

%-------------
\begin{figure}[htbp]
	\begin{center}
	\begin{tabular}{cc}
	   \subfigure[Appearances of the vacuum chamber]{
			\includegraphics[width=0.472\textwidth]{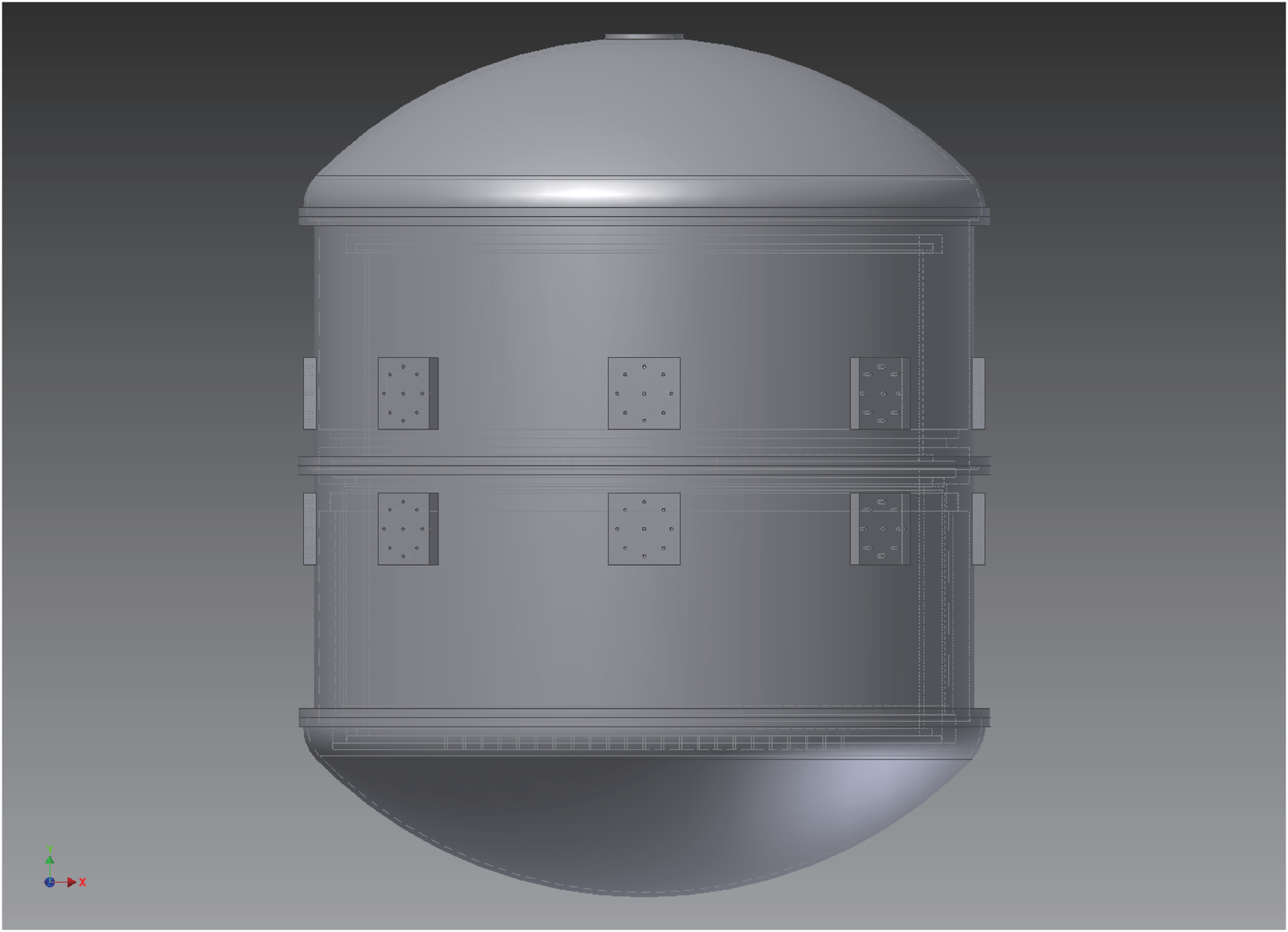}
			} &
	   \subfigure[The available space for the IRIS imager]{
			\includegraphics[width=0.47\textwidth]{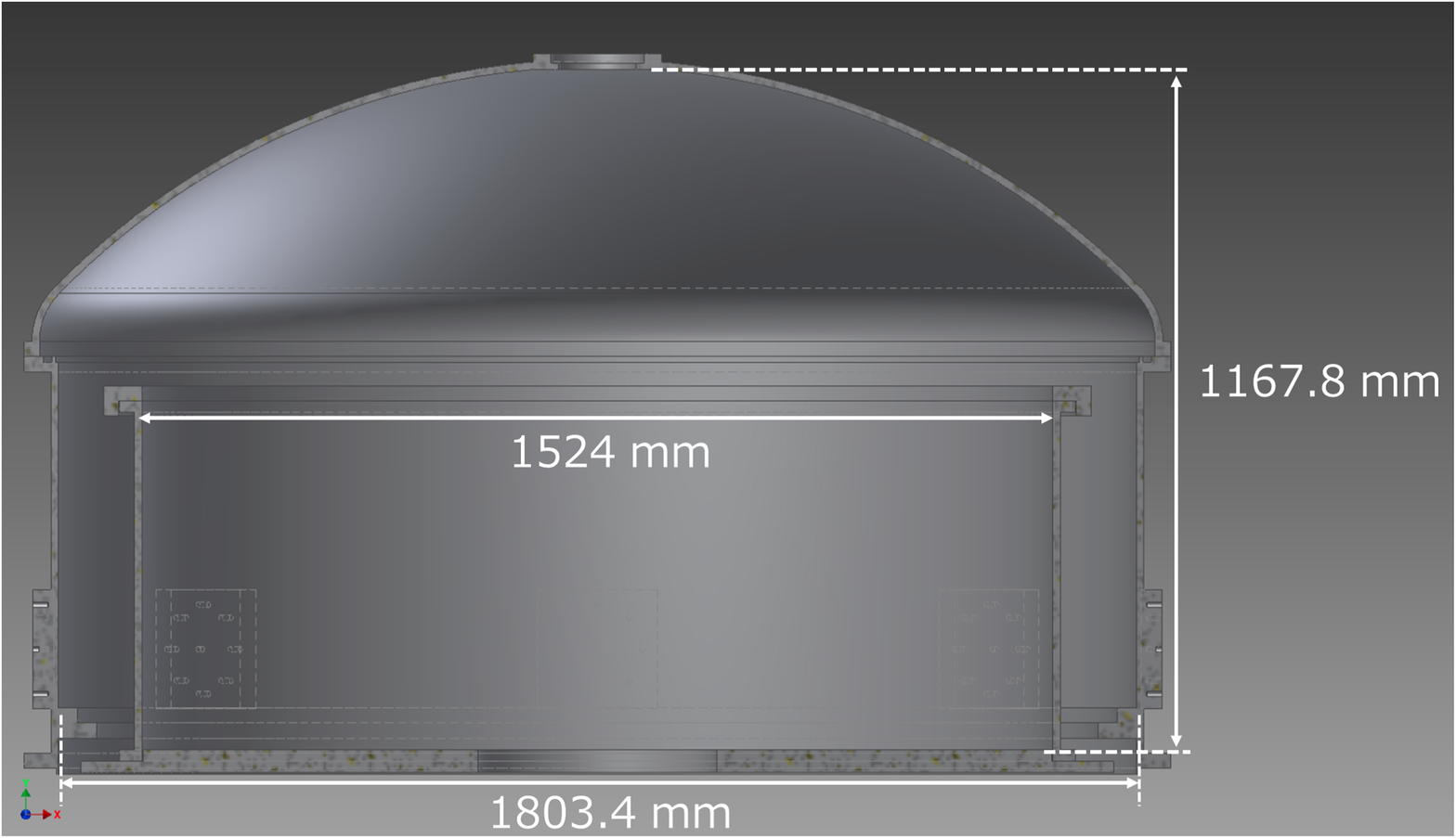}
			}
	\end{tabular}
	\end{center}
\caption{Appearances of the vacuum chamber (left) and the available space for the IRIS imager (right).
The vacuum chamber is separated into two parts.
The upper chamber is for the imager and the lower chamber is for the spectrograph.
The IRIS imager needs to be fit within the space with 1500 mm in diameter and 1150 mm in height.
Note that the size and shape of the vacuum chamber is still under discussion.}
\label{fig:IRIS_SciencePath_Dewar}
\end{figure} 
%-------------

\subsection{Effects on the Latter Optical System}
Currently, the IRIS imager optics and the IRIS spectrograph optics are continuously connected and
the part of the FoV used for imaging by IRIS imager is relayed to the IRIS spectrograph optics via a pick-off mirror.
Therefore, the parameters that will have a large impact on the latter optical system (ex. tilt angle of the image) need not to be large.

\section{OPTICAL DESIGN METHOD}
\label{sec:OPTICAL DESIGN METHOD}

Based on the design guideline described in the Section \ref{sec:OPTICAL DESIGN POLICY}, we have designed the all-reflective imaging system with the collimator and the camera. 
As a result of our study, we have developed a new type of an optical design (“Co-axis double TMA”) and incorporated it into the system.
The principle of the “Co-axis double TMA” is as shown below.

\subsection{Co-axis Double TMA}
The ”Co-axis double-TMA” is a type of optical design that eliminates the rotationally asymmetric aberration occurring at the two TMAs 
by placing the central axis of these two TMAs in parallel, resulting in n the rotationally symmetric final form of aberration.
We will further explain this type of optical design using Ritchey-Chr\'{e}tien as an example.

%-------------
\begin{figure}[htbp]
	\begin{center}
	\begin{tabular}{cc}
	   \subfigure[STEP1]{
			\includegraphics[width=0.45\textwidth]{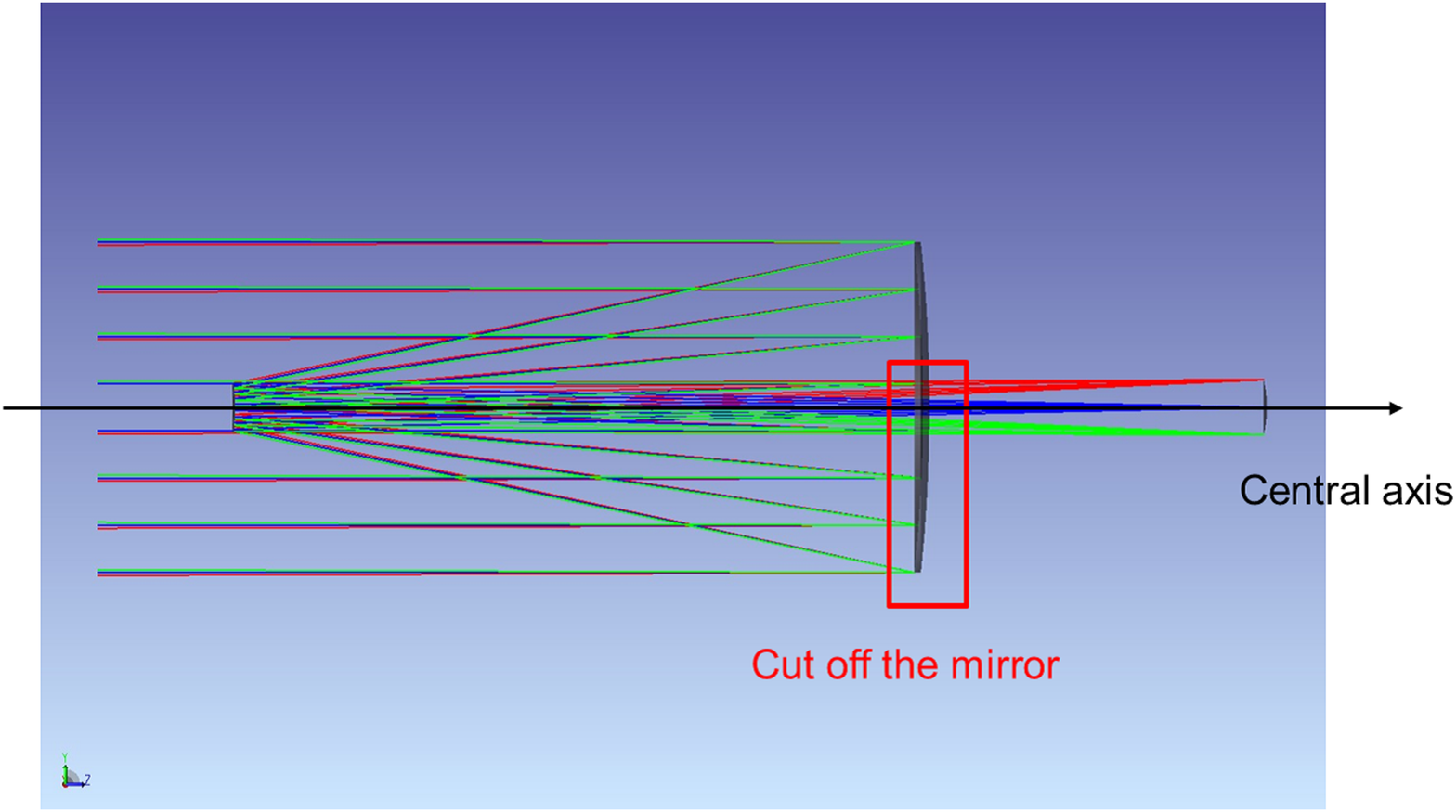}
		} &
	   \subfigure[STEP2]{
			\includegraphics[width=0.45\textwidth]{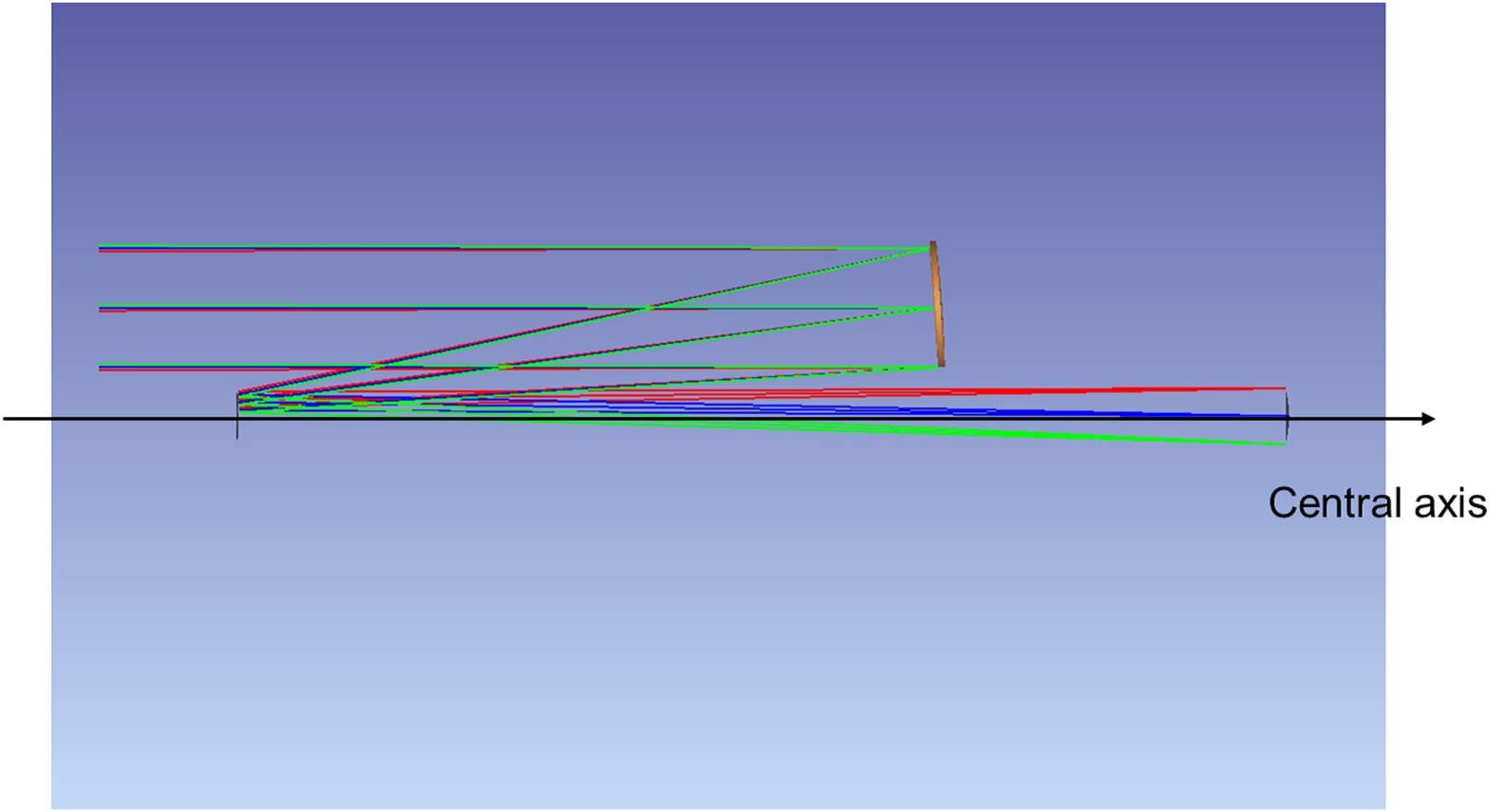}
		}\\
	 \multicolumn{2}{c}{
	   \subfigure[STEP3]{
			\includegraphics[width=0.90\textwidth]{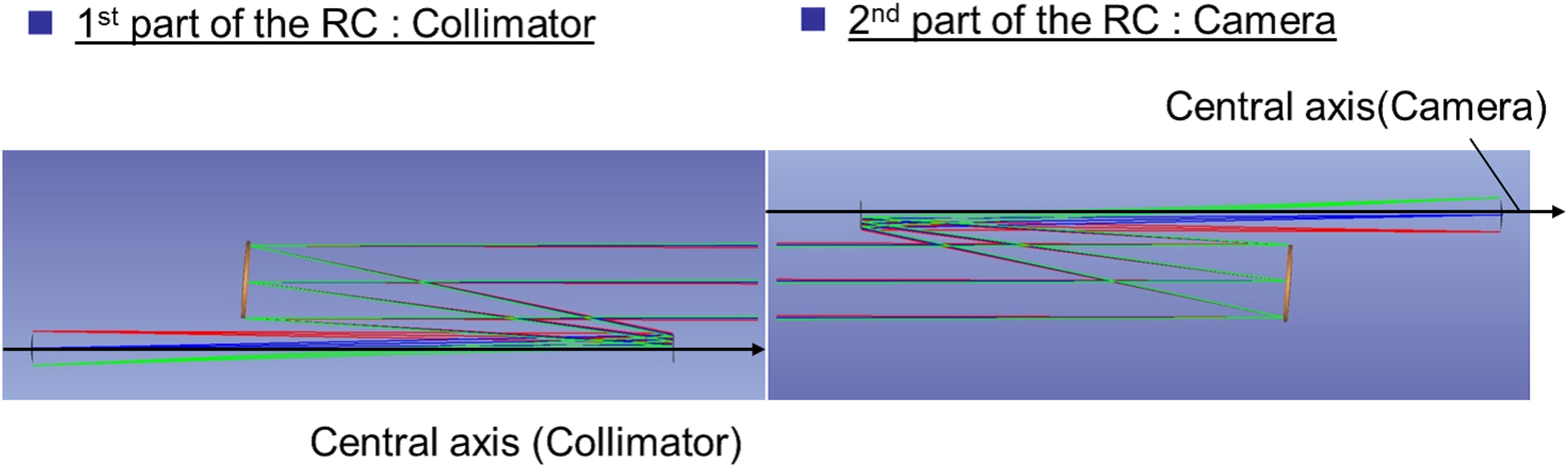}
		}
		}\\
	\end{tabular}
	\end{center}
\caption{
Conceptual Figure to explain Co-axis double TMA using Ritchey-Chr\'{e}tien telescope as an example. 
 (STEP1) : Extract part of the primary mirror of the well-designed Ritchey-Chr\'{e}tien telescope (by removing the red rectangular area).
 (STEP2) : Remaining part of Ritchey-Chr\'{e}tien telescope. 
 (STEP3) : Prepare two of the remaining part of Ritchey-Chr\'{e}tien telescope from STEP2, and reverse one of them to combine. 
}
\label{fig:SciencePathDesign_CoAxis_WTMA_RC}
\end{figure} 
%-------------

\begin{enumerate}
		\item (STEP1) : Extracting the part of the primary mirror of the Ritchey-Chr\'{e}tien telescope\\
				Well-designed Ritchey-Chr\'{e}tien telescope has features to converge the collimated light.
				Since the Ritchey-Chr\'{e}tien telescope is a rotationally symmetric optical system, its aberration is symmetrical to the central axis.
				Also, if it is sufficiently corrected within FoV, good imaging performance can be expected within the FoV.
				We will consider extracting the part of the light path of this optical system by removing the part of the primary mirror (Figure \ref{fig:SciencePathDesign_CoAxis_WTMA_RC}(a))
		\item (STEP2) : Remaining part of the Ritchey-Chr\'{e}tien telescope\\
				The remaining part of Ritchey-Chr\'{e}tien telescope after removing is no longer a rotationally symmetric optical system,
				so its aberration is rotationally asymmetric. 
				However, good imaging performance is kept since it is an extracted from the light path of the well-designed Ritchey-Chr\'{e}tien telescope (Figure \ref{fig:SciencePathDesign_CoAxis_WTMA_RC}(b)). 
		\item (STEP3) : Combining two extracted parts of Ritchey-Chr\'{e}tien telescope\\
				Prepare two of the remaining parts of the Ritchey-Chr\'{e}tien telescope from (STEP 2) and reverse one of them and combine them so that the central axis is parallel (Figure \ref{fig:SciencePathDesign_CoAxis_WTMA_RC}(c)).
				As a result, the first part of the Ritchey-Chr\'{e}tien telescope can be considered a collimator that converges the collimated light.
				On the other hand, the second part of the Ritchey-Chr\'{e}tien telescope can be considered a camera" that changes the collimated light into the converging light.
				With this arrangement, we can eliminate the rotationally asymmetric aberration occurring at the collimator and the camera, because the aberration occurring at the collimator is canceled by the aberration occurring at the camera.
\end{enumerate}

Replace the Ritchey-Chr\'{e}tien telescope with TMA (Three Mirror Anastigmat) 
and you will have the “Co-axis double TMA”.
The reason for replacing the Ritchey-Chr\'{e}tien telescope with the TMA is that the TMA is potentially enable to correct three main optical aberrations and to achieve better image quality with wide FoV. 
In fact, we were unable to sufficiently satisfy the pupil aberration requirement.
The advantages of the “Co-axis double TMA” are as follows:

\begin{enumerate}
		\item Since it can effectively eliminate the rotationally asymmetric aberration, good imaging performance can be achieved. 
				Also, it is strong with tolerances because of its aberration being only rotationally symmetric.
		\item A simple design with a small number of variables results in short calculation time for optimization and reduction of risk of falling a local minimum solution.
		\item The collimator block and the camera block are selected among the part of the well-studied reflecting  telescope. 
			    It means that the past design knowledge is available.
		\item The collimator block and the camera block are considered as a block that is complementary to each other.
		      It means that one block can be used as an effective compensator to the other.
\end{enumerate}

\section{RESULTS}
\label{sec:RESULTS}

\subsection{Nominal Performance}
We have found a promising design solution by using the “Co-axis double TMA” design method described in Section \ref{sec:OPTICAL DESIGN METHOD}. 
The following s are the performances of this design solution.

\subsubsection{Optical Layout}
\begin{figure}[htbp]
	\begin{center}
		\includegraphics[width=0.80\textwidth]{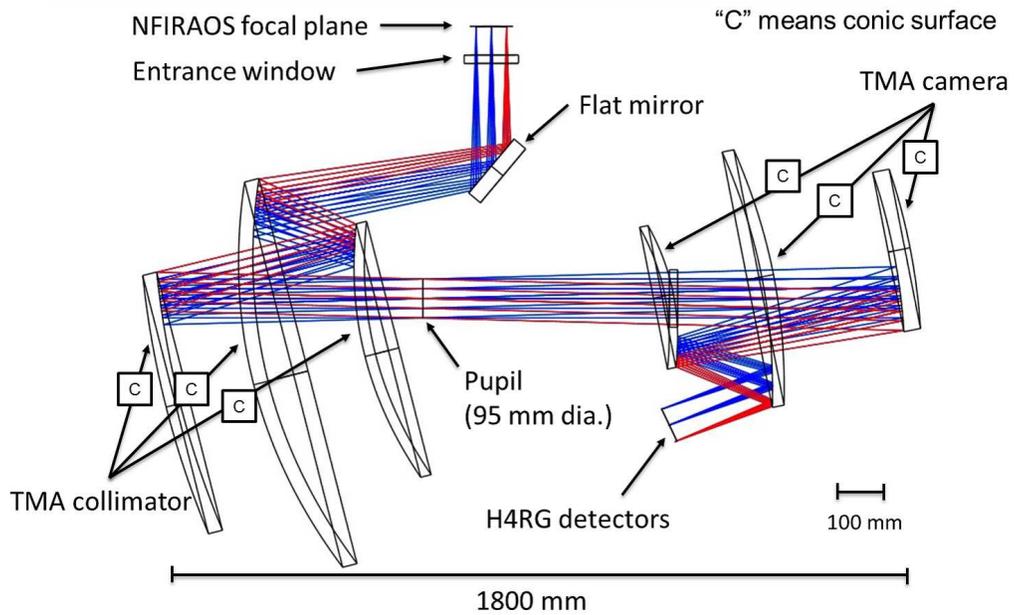}
	\end{center}
\caption{Optics layout of the latest IRIS imager design. 
			In this figure, not limited to the portions that are being used but the overall of  off-axis mirrors is described.
			The central axis of collimator/TMA and that of camera/TMA are running in parallel.			
			}
\label{fig:SciencePathDesign_CoAxis_WTMA_Layout}
\end{figure}

The basic layout of the optical system is shown in Figure \ref{fig:SciencePathDesign_CoAxis_WTMA_Layout}.
As described in Section \ref{sec:OPTICAL DESIGN METHOD}, we are using three off-axis conic mirrors as a collimator and they share the same central axis.
Similarly, three off-axis conic mirrors are used as a camera and their central axis is common as well.
And the central axis of the collimator and the camera are virtually parallel.
 
\begin{figure}[htbp]
	\begin{center}
	\begin{tabular}{cc}
	   \subfigure[Optics layout after packaging]{
			\includegraphics[width=0.55\textwidth]{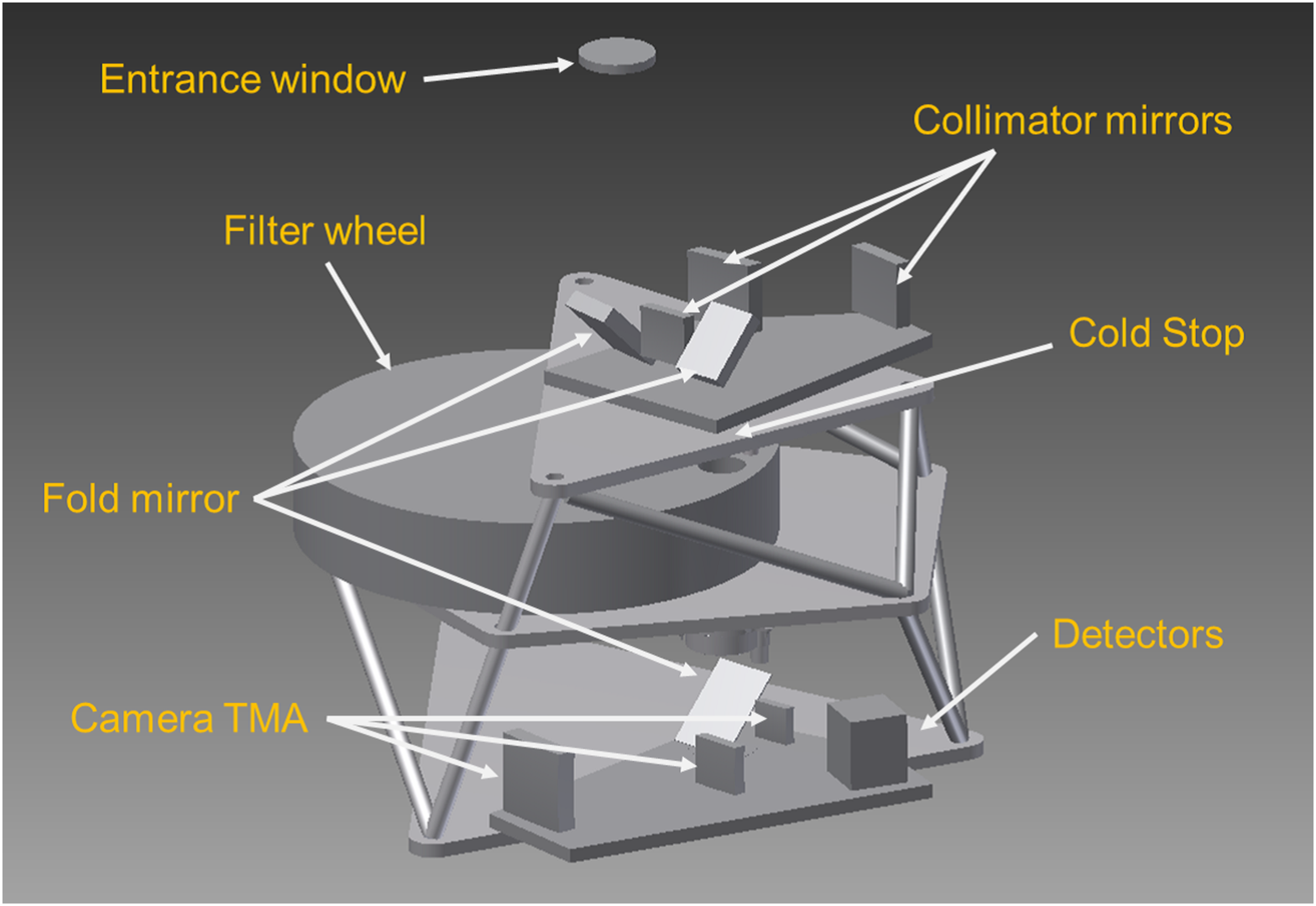}
		} &
	   \subfigure[Light path]{
			\includegraphics[width=0.37\textwidth]{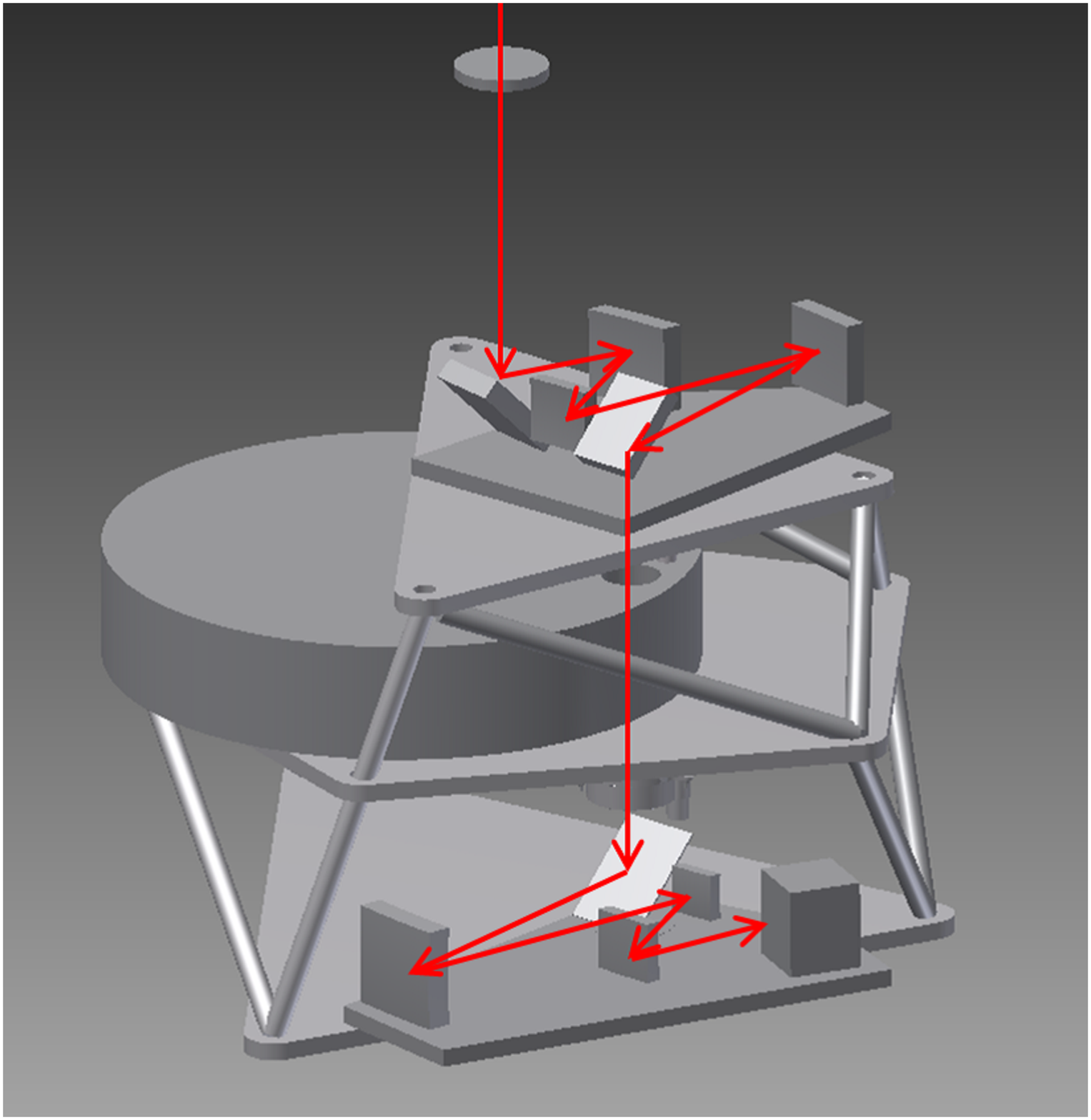}
		}\\
	\end{tabular}
	\end{center}
\caption{Layout and light path of the latest design after packaging.
			The system was able to fit within the vacuum chamber by adding three fold mirrors to six off-axis conic mirrors.}
\label{fig:SciencePathDesign_CoAxis_WTMA_PackagingLayout}
\end{figure}

Figure \ref{fig:SciencePathDesign_CoAxis_WTMA_PackagingLayout} shows the layout and light path with packaging taken into account.
By adding three fold mirrors to six off-axis conic mirrors, the optical system is made to fit in the vacuum chamber.

%-------------
%-------------

%-------------
%-------------

\subsubsection{Image Quality}
Next, we will show the imaging performance.
This optical system, with the exception of some powerless optical elements such as an entrance window, filters and ADCs, is a reflection optical system so it can be designed with the single wavelength.
Therefore we have designed the system with 1 $\mu$m single-wavelength.
Now, axial chromatic aberration caused mainly by fore optics (NFIRAOS) at each band was removed by defocusing of DMs. 

\begin{figure}[htbp]
	\begin{center}
		\includegraphics[width=0.70\textwidth]{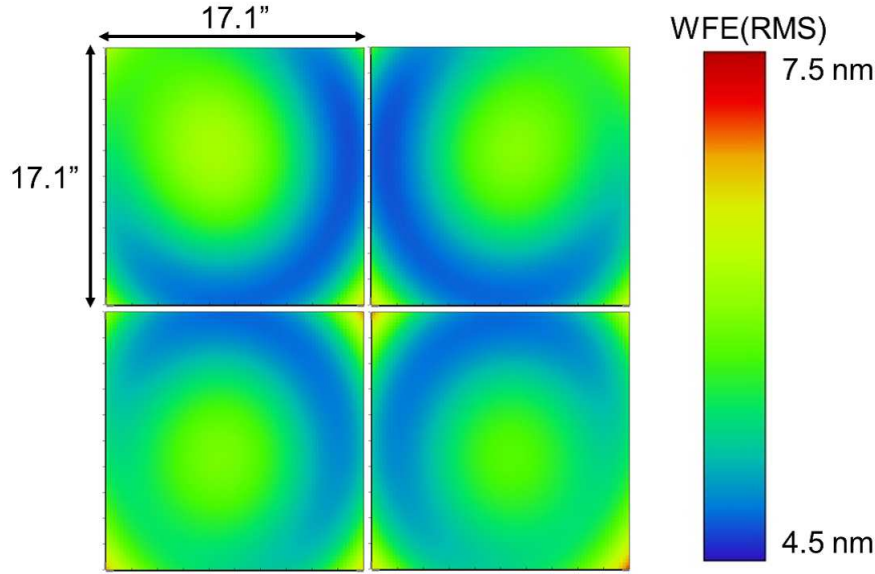}
	\end{center}
\caption{The wavefront aberration map at the overall FoV (1 $\mu$m  single-wavelength).
			You can tell that the wavefront aberration is rotationally symmetric to the center of the optical axis (center of four detectors). }
\label{fig:SciencePathDesign_CoAxis_WTMA_PerfomanceMap1}
\end{figure}

\begin{figure}[htbp]
	\begin{center}
		\includegraphics[width=0.70\textwidth]{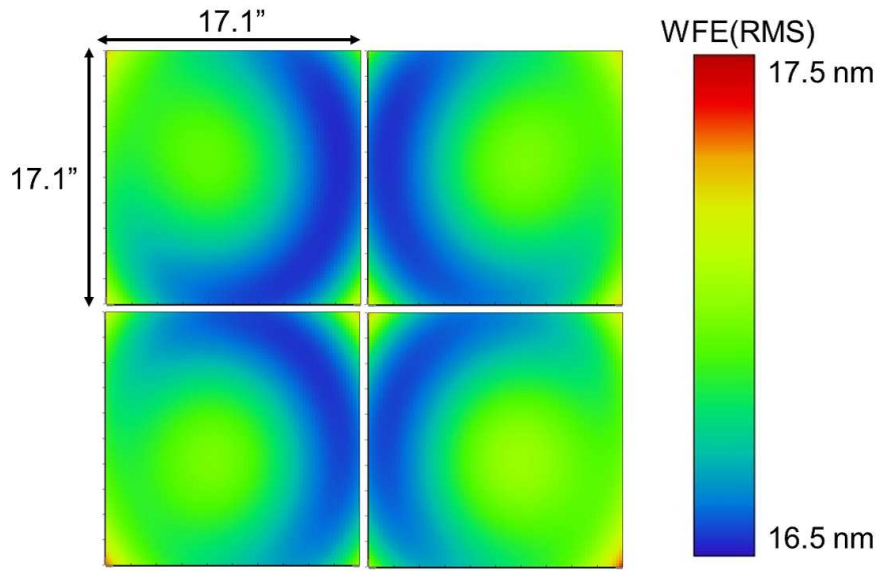}
	\end{center}
\caption{The wavefront aberration map at K-band. Since it is an almost reflective optical system, the same characteristic as that of the single-wavelength (Figure \ref{fig:SciencePathDesign_CoAxis_WTMA_PerfomanceMap1}) can be found. 
			Now, the reason of the wavefront aberration deterioration when compared to that of the single-wavelength is due to the axial chromatic aberration occurred mainly at fore optics (TMT + NFIRAOS).
			}
\label{fig:SciencePathDesign_CoAxis_WTMA_PerfomanceMap2}
\end{figure}

First, the wavefront aberration map for the overall FoV is shown in Figure \ref{fig:SciencePathDesign_CoAxis_WTMA_PerfomanceMap1} and Figure \ref{fig:SciencePathDesign_CoAxis_WTMA_PerfomanceMap2}.
Figure \ref{fig:SciencePathDesign_CoAxis_WTMA_PerfomanceMap1} shows a wavefront aberration map of 1 $\mu$m single-wavelength.
You can tell that the distribution of the wavefront aberration is rotationally symmetric to the center of the optical axis.
The reason for the distribution shown in Figure \ref{fig:SciencePathDesign_CoAxis_WTMA_PerfomanceMap1} is mainly because 
the system adjusting the field curvature with four tilted detectors cannot completely adjust the field curvature.
Except for the field curvature, the main aberration of the optical system is a spherical aberration.

Figure \ref{fig:SciencePathDesign_CoAxis_WTMA_PerfomanceMap2} shows the wavefront aberration map at K-band, which showed the worst performance among all bands.
Since this is an almost all-reflective optical system, the same distribution was found as the single-wavelength (Figure \ref{fig:SciencePathDesign_CoAxis_WTMA_PerfomanceMap1}).
Now, the reason its wavefront aberration is further deteriorated than that of the single-wavelength is because the axial chromatic aberration occurred mainly at the fore optics (TMT + NFIRAOS).
The axial chromatic aberration has the largest impact on K-band because K-band has the widest wavelength width among all bands.

Next, we will compare the results of the latest design (“Co-axis double TMA”) and the conventional design. 
The result of imaging performance comparison with the conventional design (to be called “Apt/TMA” in this paper) is shown in Figure \ref{fig:SciencePathDesign_CoAxis_WTMA_PerfomanceMap1} below.
It can be said that the “Co-axis double TMA” has the same level of imaging performance as “Apt/TMA” which utilizes transmission type optical elements. 

\begin{figure}[htbp]
	\begin{center}
	\begin{tabular}{cc}
	   \subfigure[Residual aberrations at the focal plane of the IRIS imager (TMT+NIFRAOS+IRIS)]{
			\includegraphics[width=0.65\textwidth]{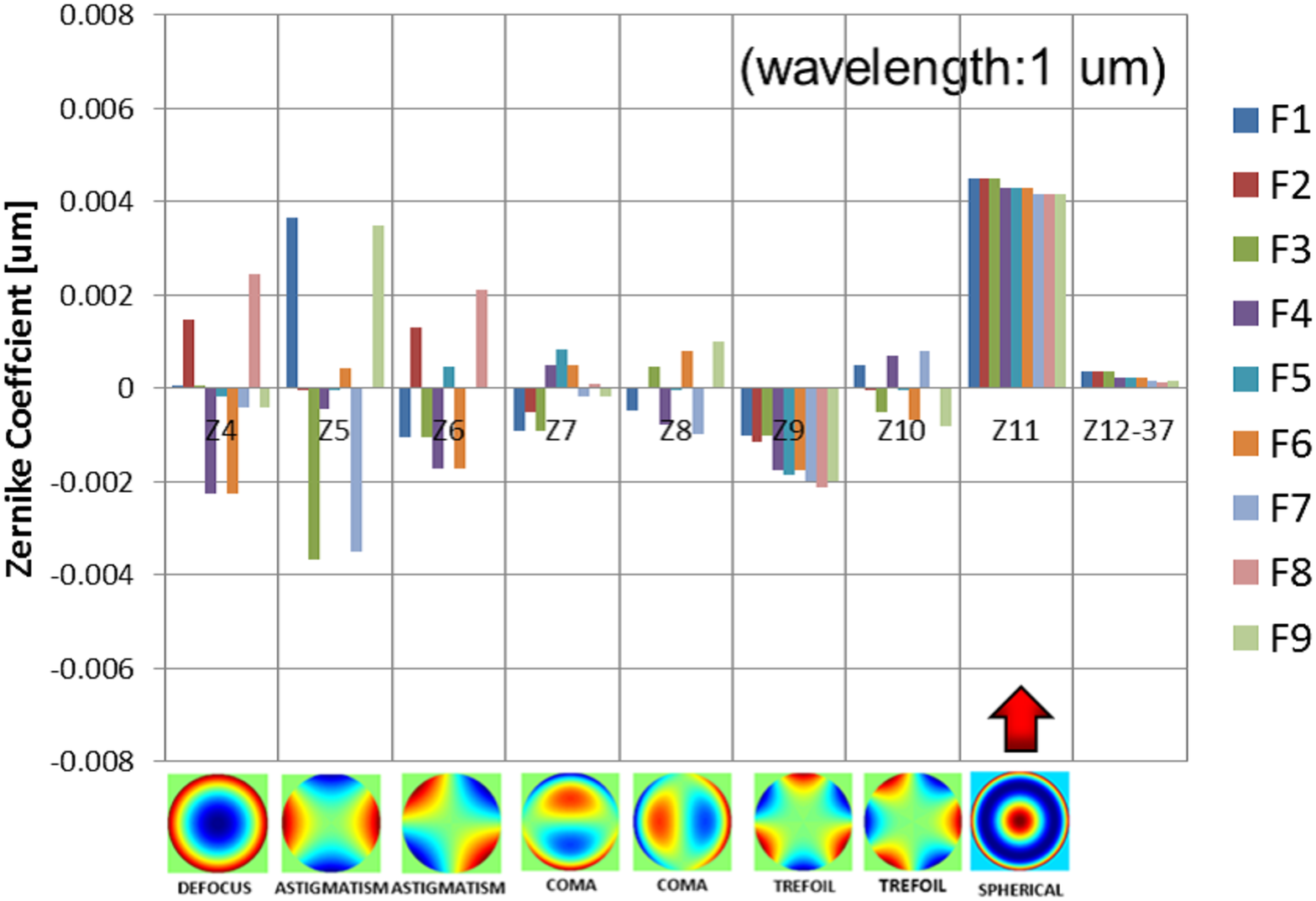}
			} &
	   \subfigure[Definition of the FoV numbers]{
			\includegraphics[width=0.25\textwidth]{image/IRIS/SciencePathDesign_FieldDefinition.eps}
			}
	\end{tabular}
	\end{center}
\caption{The residual aberration at the focal plane of IRIS (with 1 $\mu$m wavelength, eliminated the field curvature).
			The rotationally asymmetric aberrations (astigmatism (Z5, Z6) and coma aberration (Z7, Z8)), which were the dominant factors in Figure \ref{fig:IRIS_SciencePath_TMTNFIRAOS_WavefrontError}) ,
			have been significantly reduced. F1 to F9 indicates FoV numbers same as those in Figure \ref{fig:IRIS_SciencePath_TMTNFIRAOS_WavefrontError}.
			}
\label{fig:IRIS_SciencePath_TMTNFIRAOSIRIS_WavefrontError}
\end{figure}

Last, we will describe the characteristics of “Co-axis double TMA” 's aberrations.
 Figure \ref{fig:IRIS_SciencePath_TMTNFIRAOSIRIS_WavefrontError} shows the wavefront aberration at the IRIS imager focal plane (shown using Zernike coefficient).
 You can tell that the rotationally asymmetric aberrations (astigmatism (Z5, Z6) and coma aberration (Z7, Z8)) which was the dominating factors of the wavefront aberrations
at fore optics (TMT + NFIRAOS, See Figure \ref{fig:IRIS_SciencePath_TMTNFIRAOS_WavefrontError}) ) has been significantly reduced.
Instead, the spherical aberration (Z11)  that is the rotationally symmetric aberration became a dominant factor,
but this can be corrected by using DMs because this is common to all FoV.

In other words, the rotationally asymmetric aberration occurring at the front stage optical system (TMT + NFIRAOS) is eliminated by producing the opposite aberration at the IRIS itself,
making all aberrations rotationally symmetric.

%-------------
%-------------

%-------------
%-------------

%-------------
\begin{figure}[htbp]
	\begin{center}
		\includegraphics[width=0.70\textwidth]{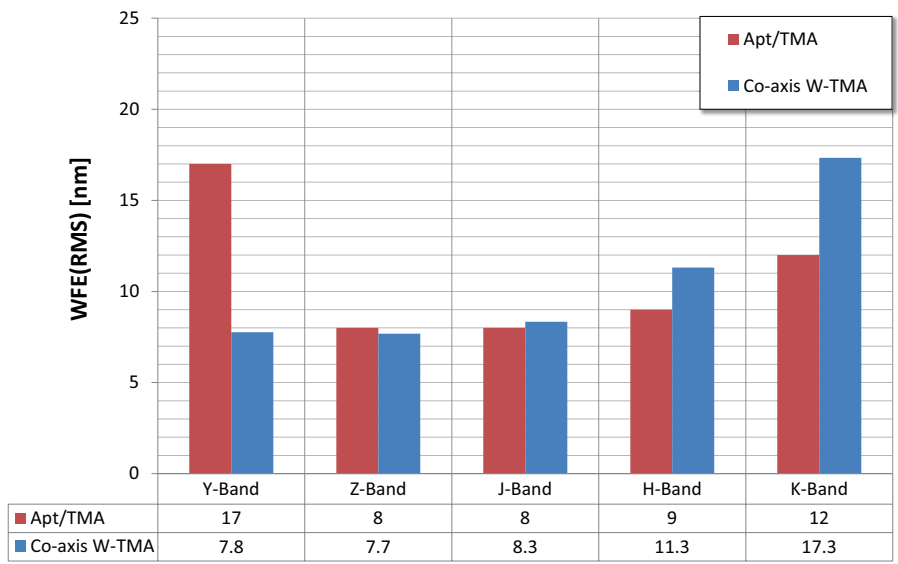}
	\end{center}
\caption{The imaging performance comparison between the latest design (“Co-axis double TMA”) and the conventional design (“Apt/TMA”).
			 “Co-axis double TMA” has the same level of imaging performance as “Apt/TMA” which utilizes transmission type optical elements.
			 Now, the reason why “Apt/TMA” and “Co-axis double TMA” show different characteristics at each band is because “Apt/TMA” eliminates the axial chromatic aberration using the transmission type optical elements.
			}
\label{fig:SciencePathDesign_CoAxis_WTMA_PerformanceComparison}
\end{figure} 
%-------------

%-------------
%-------------

\subsubsection{Throughput}
Regarding the throughput of this design, the comparison with the conventional design (“Apt/TMA”) is shown in Figure \ref{fig:SciencePathDesign_CoAxis_WTMA_ThroughputComparison} below.
We found that the “Co-axis double TMA” meets the criteria at all band ranges. Also, when compared to “Apt/TMA”, the “Co-axis double TMA” showed higher throughput by at least two percent. 
This is because the “Co-axis double TMA” is an all-reflective design and uses only three fold mirrors for packaging.

%-------------
\begin{figure}[htbp]
	\begin{center}
		\includegraphics[width=0.70\textwidth]{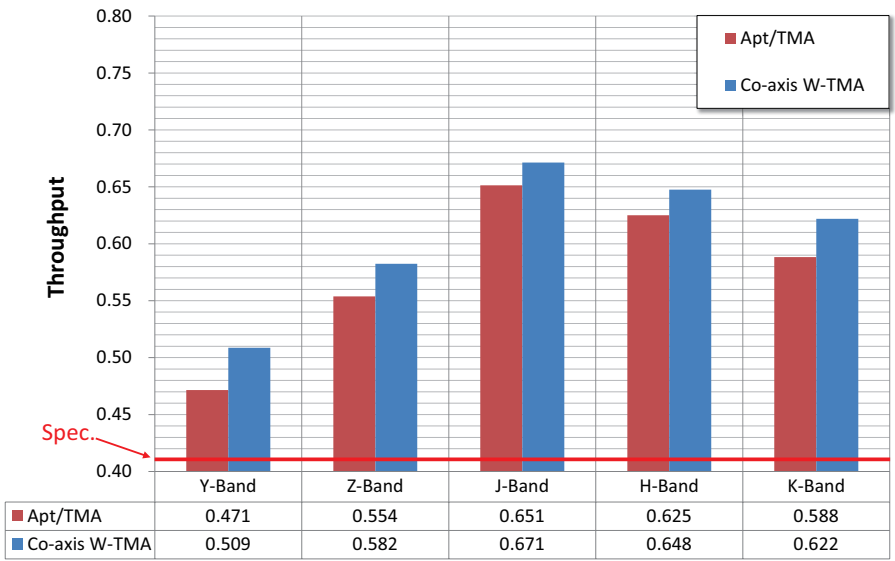}
	\end{center}
\caption{Throughput comparison between the design solution (“Co-axis double TMA”) and the conventional design solution (“Apt/TMA”). “Co-axis double TMA” is superior to “Apt/TMA” at all band range. 
			}
\label{fig:SciencePathDesign_CoAxis_WTMA_ThroughputComparison}
\end{figure} 
%-------------

\subsubsection{Distortion}
\begin{figure}[htbp]
	\begin{center}
	\begin{tabular}{cc}
	   \subfigure[Distortion volume at each band]{
			\includegraphics[width=0.40\textwidth]{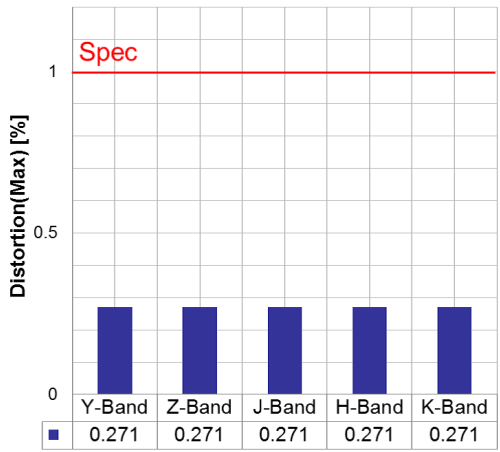}
			} &
	   \subfigure[distortion map at K-band (enlarged by 100 times)]{
			\includegraphics[width=0.40\textwidth]{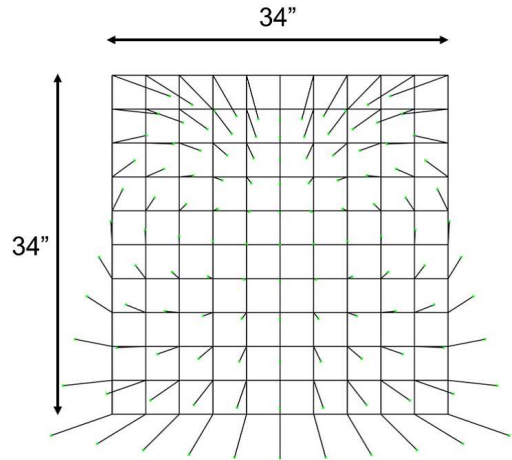}
			}
	\end{tabular}
	\end{center}
\caption{The distortion volume at each band and the distortion map at K-band for this design (enlarged by 100 times). It meets criteria at all band ranges.}
\label{fig:SciencePathDesign_CoAxis_WTMA_Distortion}
\end{figure}

The distortion volume and distortion map of this design are shown in Figure \ref{fig:SciencePathDesign_CoAxis_WTMA_Distortion}.
It meets the criteria at all band ranges.

%-------------
%-------------

\subsubsection{Pupil Aberration}
The pupil aberration and the spot diagram at the pupil position for this design solution are shown in Figure \ref{fig:SciencePathDesign_CoAxis_WTMA_Pupilabberation} below.
It meets the criteria at all band ranges.

%-------------
\begin{figure}[htbp]
	\begin{center}
	\begin{tabular}{cc}
	   \subfigure[Pupil aberration at each band]{
			\includegraphics[width=0.40\textwidth]{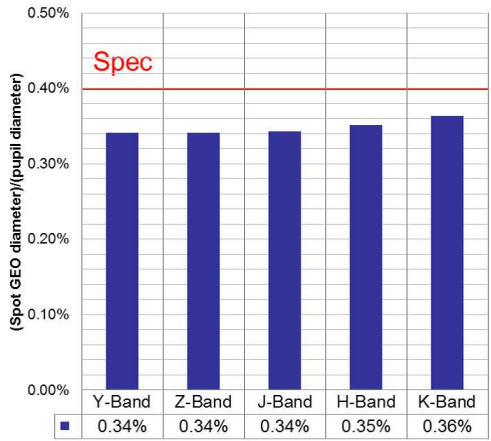}
			} &
	   \subfigure[Spot diagram at K-band]{
			\includegraphics[width=0.40\textwidth]{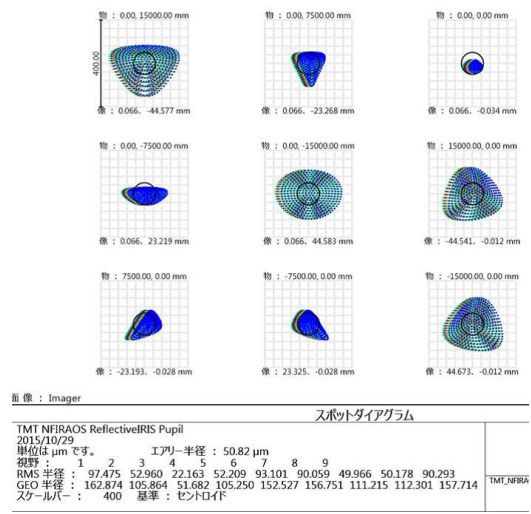}
			}
	\end{tabular}
	\end{center}
\caption{The bar graph of the pupil aberration at each band and spot diagram at K-band. It sufficiently meets criteria at all band range}
\label{fig:SciencePathDesign_CoAxis_WTMA_Pupilabberation}
\end{figure} 
%-------------

\clearpage

\subsection{Performance after Assembly Process}
This section describes the performance after the assembly process. 

\subsubsection{WFE Budget Plan}
First, the WFE budget plan determined by using the current IRIS imager is shown in Figure \ref{tab:IRIS_SciencePath_Errorbudjetplan}.
This WFE budget plan is set based on the surface irregularity error, tolerance analysis, vibration analysis, deformation effect during the cooling process, and the precision of the wavefront measurement method
such as the Phase-Diversity method.
You can tell that the surface irregularity error takes up high percentage in the WFE budget plan.

\vspace{0.01\textwidth}
%-------------
\begin{table}[htbp]
	\begin{center}
		\begin{tabular}{lr}
			\toprule[1pt]
					\textbf{Error factor} & \textbf{WFE [nm]} \\ \hline\hline
					Design Performance & 17.34 \\
					Surface Irregularity after coating & 31.75 \\
					Fabrication \& Install error & 10 \\
					Adjustment accuracy & 8 \\
					Cryo Mirrors distortion & 6 \\
					Vibration & 8 \\ 
					Measurement Error & 2 \\ \hline\hline
					Total (RSS) & 40.19 \\
			\bottomrule[1pt]
		\end{tabular}
	\vspace{0.02\textwidth}
	\caption{Current WFE budget plan of IRIS imager} 
	\label{tab:IRIS_SciencePath_Errorbudjetplan}
	\end{center}
\end{table}
%-------------

\subsubsection{Surface Irregularity Specification for Each Optical Elements}
After considering WFE budget and the machining ability of the manufacture, the surface irregularity specification of each optical element is determined as shown in table \ref{tab:IRIS_SciencePath_SurfaceIrregralitySpecification}.
Note that the surface irregularity specifications described here are the values after coating under the low-temperature environment.
These values are challenging for manufacturer mainly because these values need a firm guarantee after coating under the low-temperature environment.
 However, upon investigating manufactures, we confirmed that some are capable of meeting these challenging specifications.

\vspace{0.01\textwidth}
%-------------
\begin{table}[htbp]
	\begin{center}
		\begin{tabular}{lcc}
			\toprule[1pt]
			\textbf{Element} & \textbf{Surface irregularity}  & \textbf{WFE (RMS)}\\
				 & \textbf{after coating (RMS) [nm]}& \textbf{[nm]} \\ \hline\hline
			Window & 6 & 4.24 \\
			Filter & 6 & 4.24 \\
			Collimator 1st mirror & 6 & 12 \\
			Collimator 2nd mirror & 6 & 12 \\
			Collimator 3rd mirror & 6 & 12 \\
			Camera 1st mirror & 6 & 12 \\
			Camera 2nd mirror & 6 & 12 \\
			Camera 3rd mirror & 6 & 12 \\
			Fold mirror 1st & 3 & 6 \\
			Fold mirror 2nd & 3 & 6 \\
			Fold mirror 3rd & 3 & 6 \\\hline\hline
			Total (RSS) &  & 31.74 \\
			\bottomrule[1pt]
		\end{tabular}
	\vspace{0.02\textwidth}
	\caption{Surface irregularity specification of each element in IRIS imager.} 
	\label{tab:IRIS_SciencePath_SurfaceIrregralitySpecification}
	\end{center}
\end{table}
%-------------

\subsubsection{Tolerance Setting and Compensator Setting}
The installation and manufacturing error during an assembly process affects the imaging performance.
Thus, the appropriate tolerance and compensator that meets WFE budget need to be determined.

First, regarding the tolerance setting, after the discussion with the structure designer and the manufacturer,
it has been determined for all optical elements and optical benches as shown in table \ref{tab:IRIS_SciencePath_ToleranceValue}. 

\vspace{0.01\textwidth}
%-------------
\begin{table}[htbp]
	\begin{center}
		\begin{tabular}{lr}
			\toprule[1pt]
			\textbf{Items} & \textbf{Tolerance value} \\\hline\hline
			Shift(X,Y,Z) & 0.2 mm \\
			Tilt(X,Y) & 2 arcmin \\
			Tilt(Z) & 3 arcmin \\
			Refractive index & 0.0005 \\
			Abbe number & 0.50\% \\
			Air Thickness & 0.2 mm \\
			\bottomrule[1pt]
		\end{tabular}
	\vspace{0.02\textwidth}
	\caption{Preset tolerance values for the IRIS imager. These values are set for all optical elements and optical benches.} 
	\label{tab:IRIS_SciencePath_ToleranceValue}
	\end{center}
\end{table}
%-------------

Next, to determine the ideal compensator that meets WFE budget, we have conducted Monte Carlo simulation on the tolerance values shown above.
Setting the Monte Carlo trial number at 1,000, we have conducted two patterns of distribution; the pessimistic parabola distribution and gauss distribution.
As a result of the investigation, it was found out that the wavefront aberration specification can be met with two compensators (ex. collimator 3rd mirror and detector) 
but we could not meet the other specifications (F-number, pupil aberration, vignetting of the light flux). 
It was confirmed that an additional two compensators (ex. camera 1st mirror and ColdStop) can bring the system above criteria.
Pass rates with four compensators mentioned above are shown in table \ref{tab:IRIS_SciencePath_Passrate}.

\vspace{0.01\textwidth}
%-------------
\begin{table}[htbp]
	\begin{center}
		\begin{tabular}{lrr}
			\toprule[1pt]
			\textbf{Specification Items} & \textbf{Pass rate} & \textbf{Pass rate  } \\
			 & \textbf{(Gauss)  }& \textbf{(Parabolic) }\\ \hline\hline
			F-number & 0.992 & 0.966 \\
			Distortion & 1 & 1 \\
			WFE & 0.994 & 0.946 \\
			Pupil Aberration & 1 & 0.990 \\
			Total light transmittance & 1 & 0.989 \\ \hline\hline
			All & 0.988 & 0.925 \\
			\bottomrule[1pt]
		\end{tabular}
	\vspace{0.02\textwidth}
	\caption{Monte Carlo simulation result in case of parabolic and gauss distributions (N = 1,000). Note that these results are in the case of using four compensators. 
					Refer to table \ref{tab:IRIS_SciencePath_ToleranceValue} for the tolerance setting. }
	\label{tab:IRIS_SciencePath_Passrate}
	\end{center}
\end{table}
%-------------

\subsection{Performance in Operation}
In this section, we will describe the performance in operation.

\subsubsection{Effects of Vibration}
The main factor of the wavefront aberration deterioration caused by the vibration of the optical elements is the chief ray shift of the beam.
So we estimated the vibration value that meets the WFE budget by conducting sensitivity analysis.
As a result, the acceptable values for the most sensitive optical element were determined to be 40 nm for shift and 8 mas for tilt. 

\subsubsection{Effects of the Image Surface Rotation in Operation}
As mentioned above, the FoV of the TMT telescope rotates along with the elevation angle of the telescope.
IRIS, on the other hand, makes correction by rotating IRIS itself but this movement is equivalent to the relative rotation along the optical axis 
between fore optics (TMT + NFIRAOS) and IRIS.
This results in the rotation of the image tilt and rotationally asymmetric aberration of fore optics, which needs to be compensated. 

Upon further investigation, these effects can be sufficiently corrected by using DMs correction.

\subsubsection{Effects of ADC Rotation during Operation}
ADC rotates corresponding to the observed zenith angle, but this movement is equivalent to the rotation and position shifting of the optical axis between the collimator and the camera.
This effect is a deteriorating factor for the imaging performance as it disturbs the symmetry of the “Co-Axis double TMA”.

Upon further investigation, these effects can be sufficiently corrected by using DMs correction, too.

\section{AIV PLAN}
\begin{figure}[htbp]
	\begin{center}
		\includegraphics[width=0.90\textwidth]{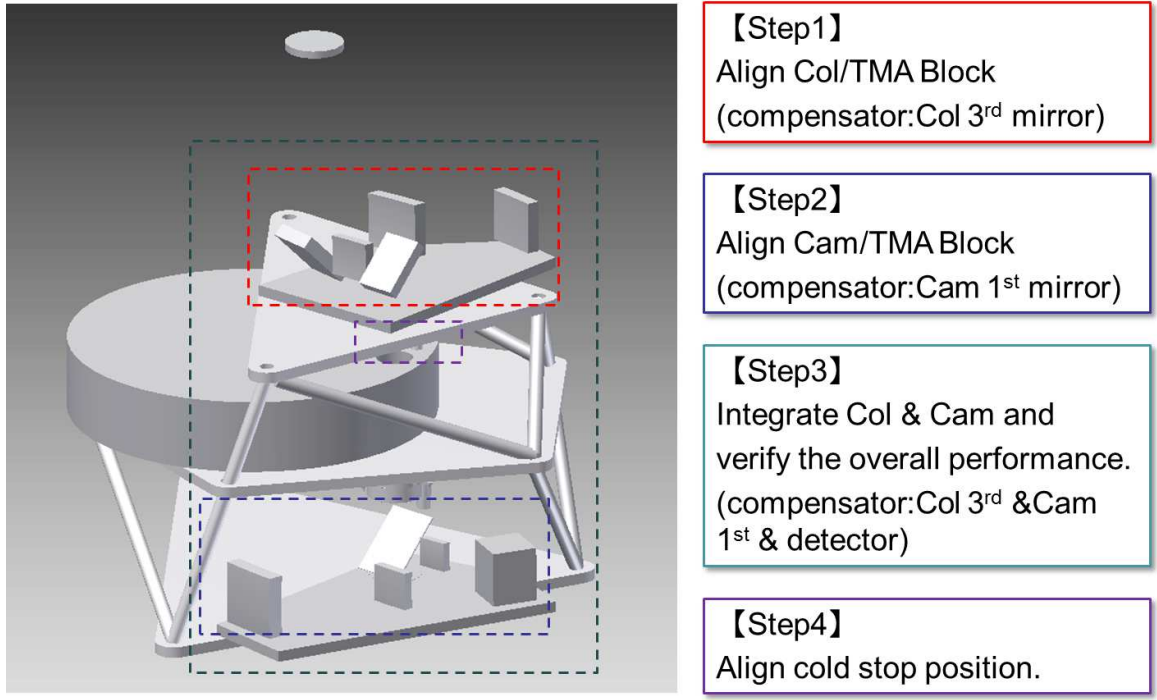}
	\end{center}
\caption{Conceptual Figure of AIV plan at the current IRIS imager}
\label{fig:IRIS_SciencePath_AIVplan}
\end{figure}

The current AIV plan based on the above-mentioned analysis is as follows: ((Refer to Figure \ref{fig:IRIS_SciencePath_AIVplan})

\begin{enumerate}
		\item (STEP1) : Installation and adjustment of collimator/TMA\\
				After installing the collimator/TMA and incident light optical system following the mechanical reference position,
				adjust collimator 3rd mirror as a compensator so that the value of the wavefront aberration or spot pattern becomes close to the design value.
				For the incident light optical system, use the collimated light or diverging light (F/15). 
		\item (STEP2) : Installation and adjustment of camera/TMA\\
				After installing the camera/TMA and incident light optical system following the mechanical reference position,
				adjust camera 1st mirror as a compensator  so that the value of the wavefront aberration or spot pattern becomes close to the design value.
				For the incident light optical system, use the collimated light or diverging light (F/17.19).
		\item (STEP3) : Combination and adjustment of collimator/TMA and camera/TMA\\
				Combine collimator and camera following the mechanical reference position and get wavefront aberration value as a combined system.
				Adjust three compensators (collimator 3rd mirror, camera 1st mirror and detectors), so that the wavefront aberration is minimized and the size of the image is correct.
				For the incident light, use diverging light (F/15) from the telescope simulator. 
		\item (STEP4) : Position adjustment of cold stop\\
				Adjust the cold stop position to the position where the diameter of the beam becomes minimal.
\end{enumerate}

%-------------
%-------------
 
\section{CONCLUSION}
We have developed a new design policy called "Co-Axis double-TMA" and have newly proposed an all-reflective optical design for the IRIS imager.
The optical system has achieved both good imaging performance and the highest throughput ever by efficiently combining the collimator/TMA and camera/TMA.
Also, we have created the error budget plan, conducted tolerance analysis and proposed an AIV plan for the design and showed that it is feasible.

\acknowledgments % equivalent to \section*{ACKNOWLEDGMENTS}       
The TMT Project gratefully acknowledges the support of the TMT collaborating institutions.  They are the California Institute of Technology, the University of California, the National Astronomical Observatory of Japan, the National Astronomical Observatories of China and their consortium partners, the Department of Science and Technology of India and their supported institutes, and the National Research Council of Canada.  This work was supported as well by the Gordon and Betty Moore Foundation, the Canada Foundation for Innovation, the Ontario Ministry of Research and Innovation, the Natural Sciences and Engineering Research Council of Canada, the British Columbia Knowledge Development Fund, the Association of Canadian Universities for Research in Astronomy (ACURA), the Association of Universities for Research in Astronomy (AURA), the U.S. National Science Foundation, the National Institutes of Natural Sciences of Japan, and the Department of Atomic Energy of India.

% References
\bibliography{report} % bibliography data in report.bib
\bibliographystyle{spiebib} % makes bibtex use spiebib.bst

\end{document}